# NEMA NU 2–2018 performance evaluation of a new generation digital 32-cm axial field-of-view Omni Legend PET-CT

Rhodri Lyn Smith[1*], Lee Bartley[2], Christopher O'Callaghan[2], Kevin M. Bradley[1], Chris Marshall[1]

[1*]The Wales Research and Diagnostic Positron Emission Tomography Imaging Centre, Cardiff University, School of Medicine, Cardiff, CF14 4XN, Wales, United Kingdom.
[2*]Radiology, Cardiff and Vale University Health Board, University Hospital of Wales, Cardiff, CF14 4XW, Wales, United Kingdom.

*Corresponding author: Rhodri Lyn Smith:
E-mail(s): SmithR50@cardiff.ac.uk;
Cardiff University, School of Medicine, PETIC, Ground Floor, C Block, University Hospital Wales Main Building, Heath Park, Cardiff, CF14 4XN

Article word count: 4987

## Abstract

A NEMA performance evaluation was conducted on the new General Electric (GE) digital Omni Legend PET-CT system with 32 cm extended field of view. This study marks the introduction of the first-ever commercially available clinical digital bismuth germanate technology. Testing was performed in accordance with the NEMA NU2-2018 standard. A comparison was made with the performance of two other commercial GE scanners with extended fields of view. Firstly, a digital lutetium yttrium orthosilicate system (the Discovery MI - 6 ring). Secondly, a non-digital bismuth germanate system (the Discovery IQ). For the Omni assessment, the tangential, radial, and axial spatial resolutions at 1 cm radial offset were measured as 3.76 mm, 3.73 mm, and 4.25 mm FWHM. The total system sensitivity to a line source at the center was 44.36 cps/kBq. The peak NECR was measured as 501 kcps at 17.8 kBq/mL. The scatter fraction at NECR peak was 35.48%, and the maximum count-rate error at and below NEC peak was 5.5%. Contrast recovery coefficients for spheres were from 52% (10 mm) to 93% (37 mm). The system does not use time of flight; thus, no assessment of timing resolution was made. The PET-CT co-registration accuracy was 2.4mm. The performance of the Omni Legend surpassed that of the Discovery MI on all NEMA tests, except for assessments of background variability (image noise). Time of flight is associated with inherent improvements in signal-to-noise ratio. In lieu of time of flight capabilities, the Omni provides software corrections in the form of a pre-trained neural network (trained on non-ToF to ToF). With such corrections, average performance is competitive when compared to ToF systems. Further validation is required to optimize clinical imaging protocols and hyperparameters associated with such software corrections and to examine the effect of non-linear corrections as target size varies, particularly for real world, clinical scans.

**Keywords:** digital PET, acceptance test, long field of view, NEMA

# 1. Introduction

Hybrid Positron Emission Tomography - Computed Tomography (PET-CT) scanners were first conceived in the early 1990s (*1*), with commercial systems first introduced in early 2001 (Discovery LS, GE Healthcare). The Discovery LS PET system consisting of 18 rings, each containing 672 bismuth germanate (BGO) scintillating crystals (4 mm × 8 mm × 30 mm crystal size) and an axial field of view (AFOV) of 152 mm was the first commercially available PET-CT system. Evolving technology in the form of both hardware and software (*2*) has resulted in increasing image quality, improved quantitative accuracy and accuracy of early disease detection from PET images. This has undoubtedly resulted in patient benefit. The purpose of this paper is to evaluate the physical performance of the new Omni Legend PET-CT system, the first commercially available clinical scanner with BGO crystals integrated with silicon photomultiplier tubes (SiPM). National Electric Manufacturer's Association/Association of Electrical Equipment and Medical Imaging Manufacturers (NEMA) performance measurements are rigorous tests performed to ensure that imaging systems are fully operational and perform according to specification. Measurements of performance are assessed before system acceptance and serve as a reference for future tests to ensure that the PET performance has not degraded over time. The standard NEMA NU 2-2018 guidelines (*3*) for PET includes a series of tests for spatial resolution, image quality, scatter fraction, count rate performance, accuracy of correction for count losses and random events, and sensitivity. We report the findings of the NEMA tests and make a comparison to other commercial GE PET-CT systems. We also present the effect of reconstruction methods on quantification using the recovery coefficient, background variability and line profiles from the NEMA image quality phantom.

# 2. Methods

## 2.1. PET-CT System

The Omni Legend PET-CT; the Revolution Maxima integrated CT component (80-140 kV) consists of 64 slices (0.625 mm thickness). The PET component has 32 cm axial FOV with LightBurst BGO crystals (4.1 mm x 4.1 mm x 30 mm) enclosed in 72 detector rings, with a total of 38016 crystals in 528 blocks, backed to 9504 silicon photomultiplier channels (SiPM). The "digital" BGO detector has the advantages of high density and stopping power, resulting in improved sensitivity. Time of Flight (ToF) capabilities are not provided by the Omni Legend; an image derived ToF correction, from non-ToF images is provided in the form of a pre-trained convolutional neural network, to enhance non-ToF images to their ToF equivalent (*4*). This is termed "precision Deep Learning (PDL)". The Omni Legend's digital BGO detector provides up-gradeability of the FOV and is designed to support future extended axial FOV upgrades. The Omni Legend is also commercially available with a 16 cm axial FOV. The system provides advanced quantitative reconstructions in the form of VUE Point HD (VPHD – 3D ordered subset expectation maximization – OSEM (*5*)) and Q.Clear (BSREM - Bayesian penalized-likelihood reconstruction (*6*)). The PDL algorithm takes as input the Q.Clear reconstructed images.

## 2.2. Measurements

Following the NEMA NU 2-2018 standard, the physical PET performance of the Omni Legend was assessed. The tests included spatial resolution, sensitivity, scatter fraction, count-rate performance, accuracy of count losses, random corrections and image quality. The impact of image reconstruction on image quality is also presented.

## 2.3. Spatial Resolution

The spatial resolution test assesses the full width half maximum (FWHM) in air of a reconstructed $^{18}F$ point source to assess the point spread function (psf). The in-air measurements represent the highest achievable resolution of the system and does not account for the effect of scatter. Furthermore, the measurements are not intended to reproduce clinical scan acquisitions, but rather to provide a standardized and reproducible measurement of reconstruction-dependent scanner spatial resolution. This may be used as a comparison of performance, over time, between scanner platforms, and across manufacturers. Three-point sources were prepared using capillary tubes, with an inside diameter of less than 1 mm and an outside diameter of less than 2 mm. Samples of $^{18}F$ with > 200MBq/ml were drawn into the capillary tube, with the fill length not exceeding 1 mm. This allows measurement of both the axial and transaxial resolutions, without rotating the source. The capillary tubes were sealed with clay and positioned with a source holder and aligned within the field of view. The sources were placed in three positions, 1 cm, 10 cm, and 20 cm vertical offsets to the center of the transverse field of view (FOV). The 1 cm position represents the center of the FOV, but positioned to avoid any possibly inconsistent results at the very center of the FOV, i.e. the 'sweet spot'. The spatial resolution was measured in two transaxial planes: one was at the center, the other was at the 1/8 position from the edge, in the axial field of view (AFOV). Image acquisition time was selected to acquire an image of > 500,000 prompt counts. The images were reconstructed using the VPHD algorithm (matrix size 384 x 384 with 6 iterations, 22 subsets, 2 mm cut off Gaussian filter). The transverse spatial resolution is the average of the radial and tangential values. Profiles across the point source response functions in all three directions (radial, tangential and axial) were generated. A parabolic fit, with linear interpolation, is fitted to the profiles; the FWHM calculated using the voxel size. The radial and tangential resolutions were averaged along the axial positions to give the transverse resolution.

## 2.4. System Sensitivity

The tomographic sensitivity determines the count rate (true coincidences) as measured by the imaging system versus the amount of activity within the FOV. A standardized source configuration is employed, comprising five concentric aluminum sleeves, each with a length of 70 cm and a thickness of 1.25 mm. These sleeves are meticulously designed with incrementally increasing diameters, ensuring precise co-axial alignment. This allows a specified distance for the positron path, before annihilation. The measurements are extrapolated to zero absorption to allow an estimate of attenuation free radioactivity to be made. A polythene tube is inserted into the aluminum sleeve sensitivity phantom which is filled with $^{18}F$ to act as a line source. The source was filled with an activity such that dead time losses are less than 5%. For the Omni this is <4MBq at imaging time. The phantom was positioned in air, supported at each end by low density materials; in this case a phantom holder at one end and an in-house designed plastic hook at the other. This set-up minimizes scatter in the center of the transaxial FOV. The phantom was centered along the z-axis of the scanner. Five, five-minute acquisitions were performed, firstly with just the smallest sheath present and subsequently increasing the wall thickness by adding the next smallest diameter outer sheath. Random coincidence events were subtracted from the prompts by using a delayed coincidence window. The count rate is corrected for decay at the start time of imaging and the attenuating sleeve material at each accumulated sleeve thickness. A function is fitted to determine the unattenuated count rate (no sleeve) $R_{corr0}$, before calculation of the system sensitivity $S_{tot} = R_{corr0}/A_{cal}$ using the activity at time of imaging $A_{cal}$. The procedure is repeated for measurements obtained when the phantom and line source are offset 10 cm from the central axis. A sensitivity profile is obtained with varying slice number along the axial offset.

## 2.5. Scatter Fraction, Count Losses, and Randoms

The scatter fraction, defined as the ratio of scatter coincidences to the sum of scattered and true coincidences, is a measure of the sensitivity of the scanner to coincidence events caused by scatter; at low count rates, random events are considered negligible. Count rate performance assesses count losses from dead time effects as a function of radioactivity. The rate of random events is also assessed. The noise equivalent count (NEC) rate is used to express the count rate performance as a function of the radioactivity concentration. The NEC estimates useful count rates of a scanner by taking into account, assuming Poisson statistics, the contribution of true events, scattered events and randoms to the total coincidence rate. Peak NEC values and the corresponding radioactivity concentration can be used as a guide to determine the optimal radioactivity to be administered to patients in a specific clinical setting. A cylindrical polyethylene scatter phantom (70 cm long and 20 cm diameter) is used as the scatter medium. The phantom has a hollow off-axis bore (45 mm radial to the center), to allow positioning of a line source consisting of a plastic tube. The line source was filled with 165MBq/ml (line source volume is 5.15ml) at imaging time. The line source was thread into the hollow bore, care taken to prevent radioactivity in the line source extending outside the boundaries of the cylindrical phantom. The scatter phantom was placed flat on the patient bed using shims and located at the center of FOV. The line source was positioned nearest the patient bed. Image frames are acquired as a decay series with sufficient counts in each frame to sample the NEC curve; this is one quarter of $T_{1/2}$ and such that each acquisition is not less than 500,000 prompt counts. The acquisition is complete when true event losses are less than 1%. The whole PET scan protocol consisted of 24 timing frames over a period of ~12 hours. PET images were reconstructed with VPHD (22 subsets, 3 iterations, 5 mm cut off Gaussian filter).

For each acquisition ($j$) (with duration $T_{acq,j}$), prompts ($C_{tot,i,j}$) and random sinograms ($C_{r,i,j}$) are generated for each slice ($i$); scattered sinograms ($C_{s,i,j}$) are estimated; allowing a determination of the true event rate by $R_{t,i,j} = C_{tot,i,j} - C_{r+s,i,j}/T_{acq,j}$, the random event rate $R_{r,i,j} = C_{r,i,j}/T_{acq,j}$ and the scatter event rate $R_{s,i,j} = C_{r+s,i,j} - C_{r,i,j}/T_{acq,j}$. The scatter fraction (SF) for each slice and acquisition is calculated by averaging across all slices, $SF = (C_{r+s} - C_r)/(C_{tot} - C_r)$. Count rate curves are calculated for total, true, random and scatter events; the NEC rate for each slice $i$ is determined by $R_{NEC,i,j} = R^2_{t,i,j}/(R_{Tot,i,j} + R_{r,i,j})$, where $R_{Tot,i,j} = C_{Tot,i,j}/T_{acq,j}$ is the total event rate. The system NEC is the sum over all slices.

## 2.6. Accuracy of Count Losses and Random Corrections

This test measures the accuracy of count losses and randoms corrections. This is performed by comparing the trues rate calculated using count losses and randoms corrections, with the trues rate extrapolated from measurements with negligible count losses and randoms.

## 2.7. Image Quality and Accuracy of Attenuation, and Scatter Correction and Quantitation

Tomographic image quality is determined by a number of different performance parameters, primarily the scanner sensitivity, tomographic uniformity, contrast and spatial resolution, and the process that is used to reconstruct the images. The purpose of the measurement is to produce images simulating those obtained in a total body imaging study. This test uses an Image Quality (IQ) phantom and a line source within a scatter medium; this scatter medium radioactivity is present outside the PET scanner field of view and mimics out-of-field radioactivity. The spheres within the image quality phantom have different diameters (10 mm, 13 mm, 17 mm, 22 mm, 28 mm, and 37 mm) with a wall thickness of 1 mm. A cylindrical insert with 5 cm outside diameter extends axially through the entire phantom; this is filled with

Styrofoam material to mimic lung attenuation. The image quality was assessed by calculating image contrast and background variability ratios for the spheres. The same experiment also estimates the accuracy of the attenuation and scatter corrections. The test is repeated three times to assess the variability in quantification. The background of the IQ phantom was filled with 5.3kBq/ml of $^{18}F$ at imaging time (the phantom volume is 9729ml), The spheres (~60ml volume in total) were filled with $21 kBq/ml$ at imaging time. This resulted in a 4:1 concentration ratio between the hot spheres and the background volume. The scatter phantom was filled with 120MBq of $^{18}F$ at imaging time. The IQ phantom was placed on the couch with the spheres facing away from the gantry. The scatter phantom was placed distal to the IQ phantom but outside the PET image FOV. The IQ phantom was centered in the sagittal and coronal planes and the axial landmark along the center of the spheres aligned. A PET-CT acquisition was performed for 7 minutes. The PET acquisition was repeated twice whilst compensating for decay and maintaining the same count statistics on subsequent acquisitions. A CT scan (120 kVp tube voltage, 115 mAs exposure, 0.95 mm pitch) was also performed for attenuation correction. The images were reconstructed with VPHD into a 384 x 384 matrix, with 6 iterations, 22 subsets and 2mm gaussian filter cut-off. For each hot sphere, a circular region-of-interest (ROI) was drawn on its central slice, encompassing the entire sphere. The average ROI counts in each of the "s" sphere ($C_s$) was calculated whilst taking into account any partial pixels. A total of 12 ROIs were drawn in the background compartment of 5 image slices (central slices ± 2), for a total of 60 ROIs; this is repeated for each sphere size. ROI positioning was automated with GE software. The average background ROI counts ($C_{Bs}$) were recorded for each sphere dimension. The contrast recovery coefficients (CRC) are defined as $CRC_s = (C_s/C_{Bs} − 1)/(conc.\ ratio − 1)$, where the conc. ratio is the calculated concentration ratio between the hot spheres and the background spheres of the same size. The background variability (BV) for each sphere "s" is also expressed as the coefficient of variation for the background spheres (i.e standard deviation for background spheres of size "s" / the average background of spheres of size "s"). Additionally, 12 ROIs in 37 mm diameters (for 5 slices) were drawn throughout the background ($C_{BL}$) and a circular ROI, 30 mm in diameter, was drawn in the center of the lung insert image on each slice (i) within the axial range ($C_{Li}$) This allows the accuracy of corrections to be assessed by the ratio of $C_{Li}/C_{BL}$ for each slice. The residual lung error (RLE) is the average of this ratio over all slices. Furthermore, the CRC, BV and RLE were assessed using reconstructions with the Q.Clear algorithm and PDL algorithm (with Q.Clear as the input). The integrated PDL algorithm can provide three levels of contrast-enhancement to noise trade off, low, medium and high. The extent of penalization when using the Q.Clear algorithm is controlled by a single parameter, β. From previous IQ phantom studies, no consistent optimum β value has been found across all sphere sizes when considering performance in terms of contrast recovery and background variability. In general, lower β values should be preferred for small structure detectability and quantification, while higher β values can be used for larger structures. A β value of 500 was used in this work which has given optimal results previously in the clinical setting for a BGO based system (Discovery IQ) (*7*) and a Digital (Discovery MI, SiPM) system (*8*). It is however noted that decreasing the β value has the effect of increasing both the contrast recovery coefficients and the background variability (*9*). The medium strength PDL algorithm was used in this work as the emphasis of the study is on the performance evaluation of the scanner, not optimization of software hyperparameters. It is an area of future work to optimize both the β parameter and strength of PDL for the Omni scanner.

## 2.8. PET-CT Co-registration accuracy

An assessment of the alignment between the PET and CT data was made to assess any co-registration error. Data was acquired with PET and CT fiducial markers at 3 locations within the PET and CT field of view at two different axial locations (20 cm from tip of table and 100 cm from tip of table). This is performed using a customized jig. IQ Spheres (17 mm, 22 mm and 28 mm) were filled with 3MBq/ml of $^{18}F$ and CT contrast medium. The spheres were attached to the jig at positions (0,1), (0,20) and (20,0) (x(cm),y(cm)) respectively. 60Kg of weight were distributed along the table in a uniform manner. Two acquisitions were performed, firstly with the phantom placed at 20 cm from the tip of the table and secondly with the phantom placed at 100 cm from the tip of the table. The centroids of the fiducial markers were calculated within the PET and CT data, and the co-registration error was determined by calculating the distance between the centroids.

# 3. Results

## 3.1. Spatial Resolution

At 1 cm radial offset the tangential, radial and axial spatial resolutions were measured as 3.73 mm, 3.76 mm and 4.25 mm FWHMs respectively. At 10 cm radial offset, the resolutions were 5.11 mm, 3.85 mm and 4.22 mm. These values along with the results at 20 cm radial offset and the full width tenth maximum (FWTM) are displayed in table 1, GE specification are displayed where available.

Table 1: Spatial Resolution

| Spatial Resolution Using VPHD | | | |
|---|---|---|---|
| | | FWHM mm | FWTM mm |
| Scan Type | Specification | Measured | Measured |
| Radial @1 cm | N/A | 3.73 | 7.71 |
| Tangential @1 cm | N/A | 3.76 | 7.69 |
| Transverse @1 cm | 4.29 | 3.75 | 7.7 |
| Axial @1 cm | 4.56 | 4.25 | 9.71 |
| Radial @10 cm | N/A | 5.11 | 9.83 |
| Tangential @10 cm | N/A | 3.85 | 7.84 |
| Transverse @10 cm | 5.06 | 4.48 | 8.84 |
| Axial @10 cm | 5.04 | 4.22 | 8.87 |
| Radial @20 cm | N/A | 7.67 | 13.77 |
| Tangential @20 cm | N/A | 4.19 | 8.11 |
| Transverse @20 cm | N/A | 5.93 | 10.94 |
| Axial @20 cm | N/A | 4.21 | 8.50 |

## 3.2. System Sensitivity

The total system sensitivity with the line source at the center of the FOV was 44.36 cps/kBq. At 10 cm off-center, this result was 44.63 cps/kBq. The axial sensitivity profile at 10 cm offset is shown in figure 1. Table 2 shows the result in comparison to the GE specification.

Table 2: System Sensitivity

| Sensitivity | | | |
|---|---|---|---|
| Sensitivity | Units | Specification | Measured |
| Center | cps/kBq | 41.4 | 44.36 |
| 10 cm off center | cps/kBq | 38.7 | 44.63 |

## 3.3. Scatter Fraction, Count Losses, and Randoms

The count rates of NEC, prompts, trues, randoms, and scatter varied with radioactivity concentration and are plotted as curves shown in figure 2. The peak NEC rate is measured as 501 kcps at 17.8 kBq/mL activity concentration. The scatter fraction at Peak NEC was 35.48%. The max/min count rate relative errors at NECR peak are 5.5%. These values are compared against GE specification in tables 3 and 4 respectively.

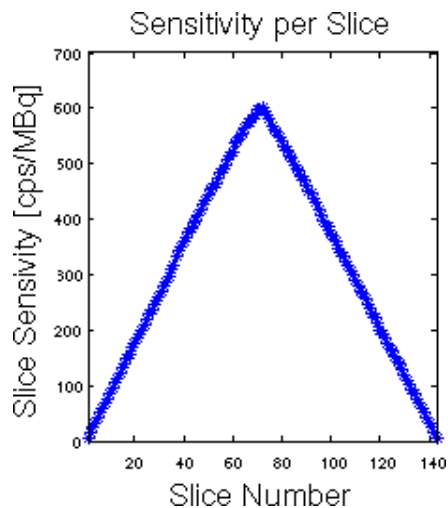

**Fig. 1**: Axial sensitivity profile at the 10 cm offset profile

**Table 3**: NECR

| | Noise Equivalent Count Rate | | |
|---|---|---|---|
| | Units | Specification | Measured |
| NECR Peak Lower Limit | kcps | 450 | 501 |

**Table 4**: Count Rate Accuracy

| | Noise Equivalent Count Rate | | |
|---|---|---|---|
| | Units | Specification | Measured |
| Max absolute error (below peak NECR) | % | 5.5 | 4.4 |

# 4. Image Quality

The 4:1 averaged ratio of sphere to background, CRC ranged from 53% (10 mm) to 93% (37 mm) for the hot spheres (averaged from the three acquisitions). The lung residual was measured to be 11%. All CRCs and corresponding BV along with GE specification are summarized in Table 5; this is presented for VPHD reconstruction, Q.Clear and the result of the PDL processing. The central slice of the image quality phantom for the 4:1 measurement, the CRC curves, BV and lung residual, for a single VPHD reconstructions are shown in figure 4. Figure 5 displays the central slice of the image quality phantom for reconstructions using VPHD, Q.Clear and PDL. Also presented in Figure 6 are the horizontal and vertical line profiles for the spheres within the central slice; displayed are line profiles for reconstructions with VPHD, Q.Clear and PDL.

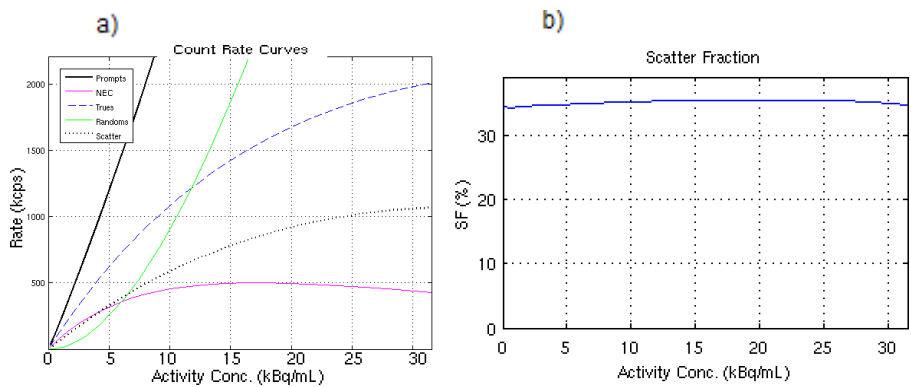

**Fig. 2**: (a) Measured count-rate curves of prompt, delayed, scatter, true, and NEC rates, (b) scatter fraction curve versus activity concentration.

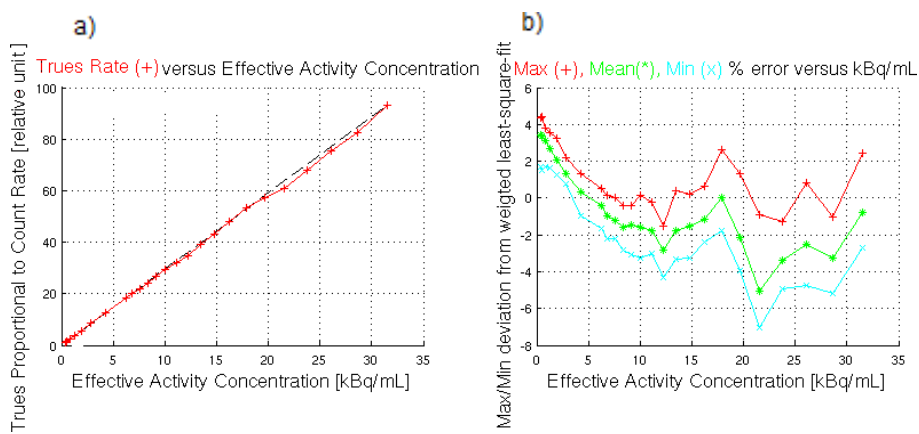

**Fig. 3**: (a) relative true rates versus effective activity concentration. (b) the maximum and minimum relative count-rate error curves for difference activity radio concentration.

## 4.1. PET-CT Co-Registration

The Maximum PET-CT Co-registration error was measured to be 2.4 mm. Table 6 displays the result together with the GE specification.

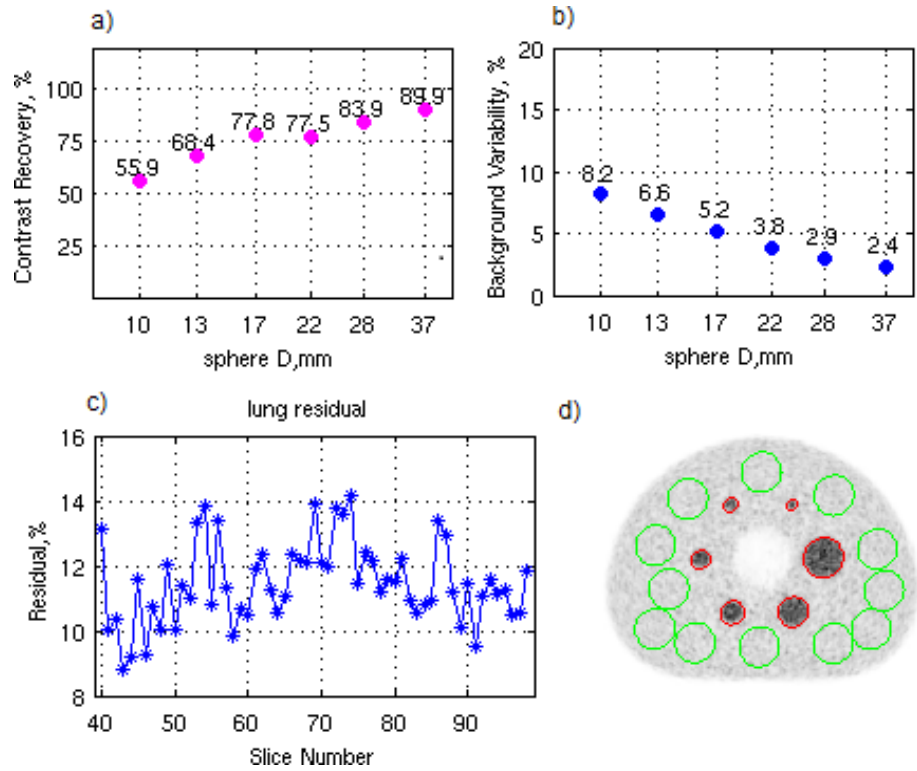

**Fig. 4**: (a) Contrast recovery curves from a NEMA IQ acquisition reconstructed using VPHD, (b) displays the background variability, (c), the lung error versus slice number and (d) the central reconstructed slice showing region of interest positions.

## 5. Discussion

In this work the PET performance of the Omni Legend system with a 32-cm axial field-of-view was evaluated. The Omni Legend provides the first digital BGO based PET-CT system on the market. The emergence of digital PET-CT has seen a leap forward in PET image quality and lesion detectability, particularly for smaller lesions (*10*). Increasing the scanner geometry, with an extended axial field-of-view, inherently increases the system sensitivity, albeit at increased cost. The BGO crystal, at a lower cost than L(Y)SO boasts higher detection efficiency and plays an important role in selecting a suitable scintillator for extended FOV's. Further advantages may be obtained by using a BGO coupled to digital detector technology. BGO does however lack the timing resolution of Lu-based scintillators. Promising work does however exist in using BGO Cherenkov based emission for improved coincidence timing. (*11*). As the Omni Legend is a digital BGO system with extended FOV, for comparative purposes, we assess its measured performance against two commercially available GE systems. Firstly, the 6 ring

Discovery MI, which has an axial FOV of 30 cm consisting of lutetium-yttrium-oxyorthosilicate (LYSO) scintillator crystals (3.95 mm × 5.3 mm × 25 mm) backed onto SiPM arrays. Secondly, at the time of writing, the Discovery IQ, which from the GE BGO range has the highest reported axial FOV of 26 cm (other than the Omni Legend) (*12*). The Discovery IQ crystals (6.3 mm × 6.3 mm × 30 mm) are coupled with traditional photo-multiplier tube PMT technology. The summary comparison is described in table 7. It is clear to observe that the sensitivity of the Omni, far surpasses the Discovery MI and Discovery IQ by ~30% and ~65% respectively. The latter increase being the result of the SiPM detectors versus the PMT, which is also reflected by the increase in contrast recovery of the Omni versus the Discovery IQ. Time of flight reconstructions have negligible impact on spatial resolution of reconstructed images at current timing resolutions (*13*). Spatial resolution is therefore compared with VPHD for all three systems. With the larger voxel size, the Discovery IQ's spatial resolution as expected is inferior. Taking the average spatial resolution (Radial, tangential, axial) at all offsets (1 cm, 10 cm and 20 cm), the Omni resolution is 4.55 mm versus 4.59 mm of the Discovery MI. Time of flight capabilities shows demonstrable improvements in Signal to Noise ratio (SNR) and hence greater contrast to noise ratio (*14*). Considering the image quality results. A comparison of VPFX (TOF-OSEM) from the Discovery MI 6-ring to the VPHD of the Omni Legend, we observe an average improvement (average across all spheres) in contrast recovery of 4% for the Omni. The Omni does, however, demonstrate an average increase in background variability by 24% in comparison to the Discovery MI. The background variability reflects the noise in the image and demonstrates the improved performance of VPFX versus VPHD. The Omni also exhibits an 18% increase in background variability in comparison to the Discovery IQ. The Omni does have a smaller voxel size in comparison to the Discovery IQ which could account for this reduced performance. Increased background variability in SiPM versus PMT systems has however previously been reported (*15*) with benefits of digital systems over analogue mainly being the benefit advanced reconstruction algorithms (*16*).

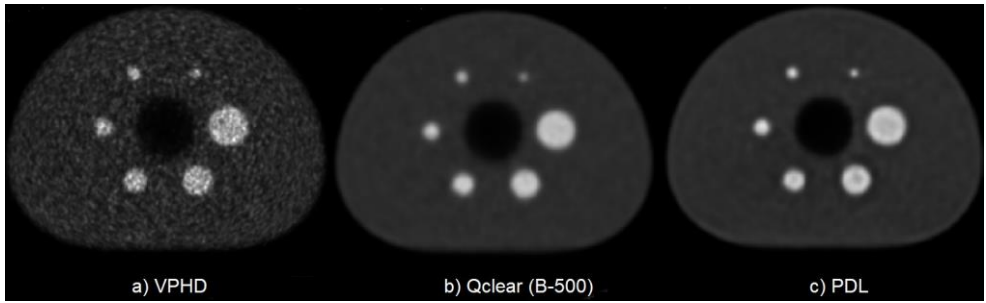

**Fig. 5**: Central slice of the IQ phantom with images reconstructed using (a) % VPHD(b) Q clear with a *β* parameter of 500 and (c) the Q clear image processed with the PDL network. All images are displayed with a grey linear color scale with maximum cut-off of, 250000 Bq/ml.

**Table 5**: Image Quality

| Image Quality | | | | | | | |
|---|---|---|---|---|---|---|---|
| | Hot spheres | | | | | | Lung Error |
| Diameter | 10 mm | 13 mm | 17 mm | 22 mm | 28 mm | 37 mm | 50 mm |
| Specified Contrast % | 30 | 40 | 50 | 60 | 60 | 60 | 19 |
| Measured Contrast VPHD % | 52.7 | 63.7 | 76.7 | 78.8 | 84.9 | 92.5 | 11.4 |
| Specified Background % | 12 | 10 | 9 | 7 | 6 | 5 | - |
| Measured Background VPHD% | 8.7 | 6.8 | 5.3 | 4.2 | 3.1 | 2.5 | n/a |
| Measured Contrast (500) Qclear% | 32.4 | 52.5 | 67.9 | 73.1 | 85.8 | 92.5 | 6.8 |
| Measured Background (500) Qclear % | 2.1 | 1.8 | 1.6 | 1.4 | 1.2 | 1 | n/a |
| Measured Contrast (500) PDL % | 48.5 | 62.3 | 78.6 | 74.4 | 89.1 | 96.5 | 2.6 |
| Measured Background (500) PDL% | 2.9 | 2.6 | 2.4 | 2.4 | 2.3 | 2.2 | n/a |
| **Note:** The measured contrast should be greater than or equal to the specified values. The measured lung error should be less than or equal to the specified values. The measured background values should be less than or equal to the specified values. | | | | | | | |

**Table 6**: PET-CT Co-Registration

| PET-CT Co-Registration | | | |
|---|---|---|---|
| Scan Type | Units | Specification | Measured |
| Max Co-registration Error | % | 5 | 2.4 |

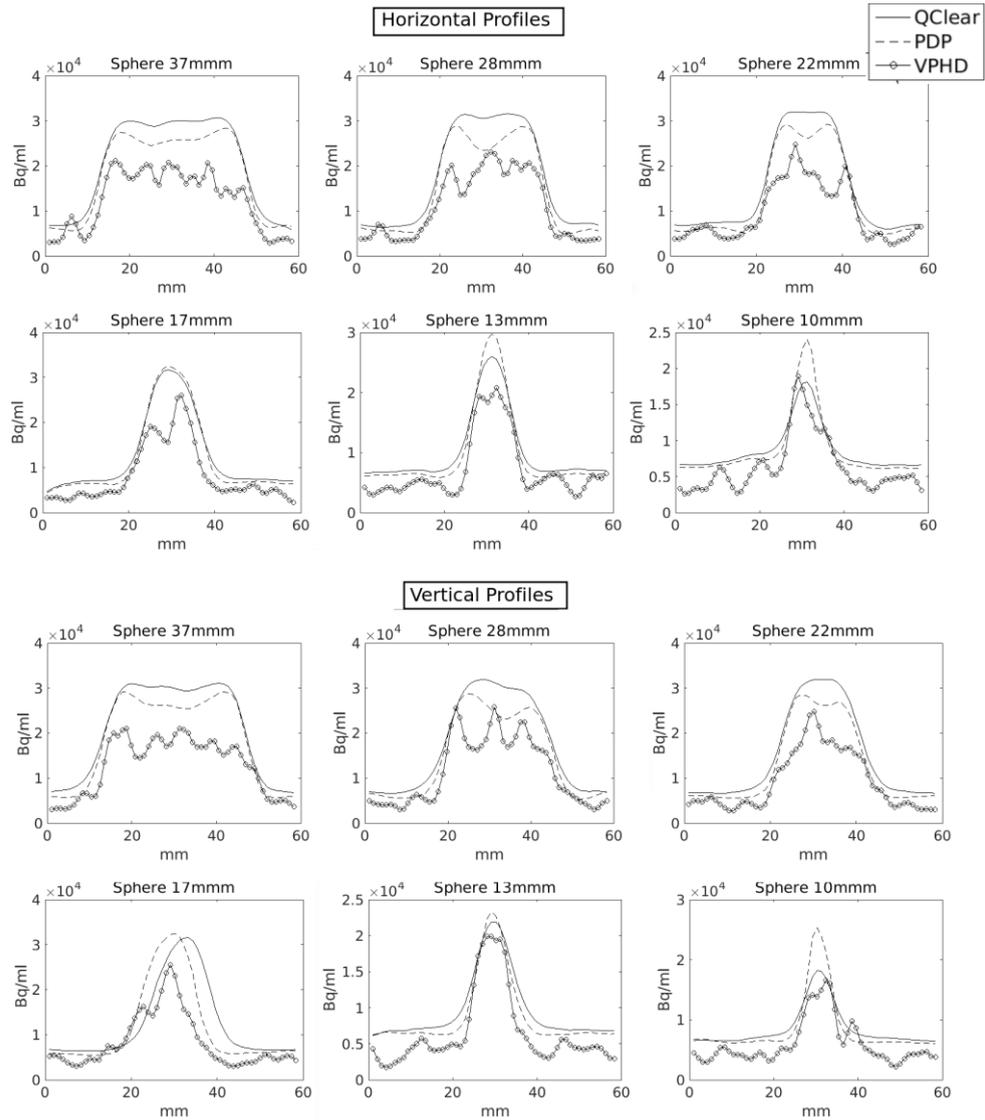

**Fig. 6**: Horizontal and vertical profiles taken from the spheres within the central slice of the NEMA IQ phantom. Displayed are the profiles when images are reconstructed with VPHD, Q clear, with a *β* parameter of 500 and post processed with PDL.

The Omni Legend includes the new precision deep learning (DL) technology (4) with the intent to provide the image quality performance benefits most associated with hardware-based Time-of-Flight. Comparing contrast recovery for VPHD, Q.Clear (*β* = 500) and PDL; the Omni demonstrates when averaged across all spheres, a 4% increase, 7% reduction and 4% increase in comparison to the VPFX of the Discovery MI respectively. The background variability demonstrates a 126% increase, 61% reduction and 36% reduction in comparison to the Discovery MI (VPFX), when utilizing VPHD, Q.Clear and PDL respectively. Interestingly, PDL increases background variability in comparison to Q.Clear. Nonetheless, the advanced algorithms reduce background variability to superior levels in comparison to the VPHX of Discovery MI. This demonstrates, in average terms, the Omni's ability to recover performance comparable to a ToF system. It can be observed in figure 6 that count recovery from spheres is more pronounced with Q.Clear in comparison to VPHD and DL for spheres (simulated lesions) larger than 17 mm. For smaller lesions, the count recovery for PDL exceeds Q. Clear ($\beta$=500) significantly. For the 10 mm lesion, Q.Clear in comparison to VPHD, increases counts, when averaged along the horizontal and vertical profiles by 3%; for PDL, this value is 40%. This is as expected for a BSREM reconstruction and will vary with the $\beta$ value chosen, with greater contrast recovery for lower $\beta$ values due to less noise penalization and conversely, less contrast recovery for higher $\beta$ values due to greater noise penalization. Furthermore, the PDL algorithm, in comparison to Q.clear, provides relatively less contrast recovery for large spheres and greater contrast recovery for small spheres (Table 5). This is as expected, since PDL was trained to mimic time-of-flight like performance. It is also important to understand that the PDL reconstruction algorithm was trained using clinical PET-CT scans from a variety of PET-CT scanners and not phantoms. Therefore, using standard phantoms to measure PDL performance should only be considered as a guide to performance. Variabilities inherently exist in the optimal value of $\beta$ across lesion volumes when utilizing Q.Clear (*17*) which serves as the input to PDL. Future work in validating performance with varying hyperparameters, both the Q.Clear $\beta$ parameter and strength of PDL is required. This will likely demonstrate improvements in both contrast recovery and background variability.

## 6. Conclusion

A NEMA performance evaluation is made of the new digital Omni Legend. A comparison was made to the Discovery MI (6 ring) and Discovery IQ. The Omni demonstrates increased sensitivity in comparison to its counterparts, allowing the possibility of shorter scan times and / or less patient dose. Omni image quality in terms of spatial resolution and contrast recovery is competitive even in comparison to an LYSO system (Discovery MI). The Omni does, however, present increased background variability (noise) in comparison to the Discovery MI and Discovery IQ. Software corrections are included (Q.Clear, PDL) that attempt to circumnavigate the lack of ToF capabilities. In phantom studies, these "on average" demonstrate themselves to be successful in terms of reducing background variability. Further validation work is required to assess the effect of hyperparameters on such corrections, particularly in the presence of varying geometry (e.g lesion size / attenuating medium). Further improvements in image quality are likely following this optimization. This should be explored in combination with optimizing patient dose and imaging time.

**Table 7**: Comparison

| System | | Omni Legend | Discovery MI 6 ring | Discovery IQ |
|---|---|---|---|---|
| Assessment | | Measured NEMA NU2-2018 | Zeimpekis et al.(*18*) NEMA NU2-2018 | Reynés-Llompart et al (*9*). NEMA NU2-2012 |
| Parameter ↓ | | | | |
| Axial FOV (cm) | | 32 | 30 | 26 |
| Detector (type) | | SiPM | SiPM | PMT |
| Scintillator type | | BGO | LYSO | BGO |
| Scintillator size (mm) | | .4.1 ×4.1 × 30 | 3.95 ×5.3 × 25 | 6.3 ×6.3 × 30 |
| Sensitivity (cps/kBq) (center) | | 44.63 | 32.64 | 22.8 |
| Sensitivity (cps/kBq) (10cm) | | 44.36 | 32.88 | 20.43 |
| Peak NECR (kcps) | | 501 | 434.3 | 123.6 |
| Peak NECR conc. (KBq/ml) | | 17.8 | 23.6 | 9.1 |
| SF % at Peak NECR | | 35.48 | 40.21 | 36.2 |
| Max error at peak NECR (%) | | 5.5 | 3.95 | 3.9 |
| Spatial Resolution | | VPHD | VPHD | VPHD |
| Radial @1cm | | 3.73 | 3.72 | 4.5 |
| Tangential @1cm | | 3.76 | 3.87 | 4.7 |
| Transverse @1cm | | 3.75 | 3.8 | 4.6 |
| Axial @1cm | | 4.25 | 4.26 | 4.8 |
| Radial @10 cm | | 5.11 | 4.8 | 5.6 |
| Tangential @10 cm | | 3.85 | 3.79 | 5.1 |
| Transverse @10 cm | | 4.48 | 4.3 | 5.4 |
| Axial @10 cm | | 4.22 | 4.55 | 4.8 |
| Radial @20 cm | | 7.67 | 7.63 | 8.5 |
| Tangential @20 cm | | 4.19 | 4.21 | 5.5 |
| Transverse @20 cm | | 5.93 | 5.95 | 7 |
| Axial @20 cm | | 4.21 | 4.50 | 4.8 |
| Image Quality | | VPHD | VPFX | VPHD |
| Diameter Hot Sphere | | | | |
| 10mm | % Contrast | 52.7 | 54.5 | 25 |
| 13mm | % Contrast | 63.7 | 63.2 | 40 |

| 17 mm | % Contrast | 76.3 | 68 | 61 |
| 22 mm | % Contrast | 78.9 | 76.9 | 68 |
| 28 mm | % Contrast | 84.9 | 82.4 | 64 |
| 37 mm | % Contrast | 92.5 | 85.8 | 68 |
| Lung error | | | | |
| 50 mm | % Contrast | 11.4 | 3.16 | 22.2 |
| Background Variability | | | | |
| 10 mm | % | 8.7 | 6.8 | 5.5 |
| 13 mm | % | 6.8 | 5.0 | 4.9 |
| 17 mm | % | 5.3 | 4.0 | 4.2 |
| 22 mm | % | 4.2 | 3.2 | 3.6 |
| 28 mm | % | 3.1 | 2.5 | 3.4 |
| 37 mm | % | 2.5 | 1.9 | 3.3 |

## Disclosure

There are no conflicts of interest to declare relevant to this article.